\documentclass[aps,showpacs,preprintnumbers,amsmath,epsf,amssymb,epsfig,nofootinbib]{revtex4}
\usepackage{bm}
\begin{document}

\title{Stochastic approach of gravitational waves in presence of a
decaying cosmological parameter from a 5D vacuum}
\author{$^{1}$Silvina Paola Gomez Mart\'{\i}nez \footnote{
E-mail address: gomezmar@mdp.edu.ar} , $^{3}$ Lucio F. P. da Silva
\footnote{ E-mail address: luciofabio@fisica.ufpb.br}  ,
$^{3}$Jos\'e Edgar Madriz Aguilar\footnote{ E-mail address:
jemadriz@fisica.ufpb.br} and $^{1,2}$Mauricio
Bellini\footnote{E-mail address: mbellini@mdp.edu.ar}}

\address{$^1$ Departamento de
F\'{\i}sica, Facultad de Ciencias Exactas y Naturales, Universidad
Nacional de Mar del Plata, Funes 3350, (7600) Mar del Plata,
Argentina.\\
$^2$ Consejo Nacional de Investigaciones
Cient\'{\i}ficas y T\'ecnicas (CONICET). \\
$^3$ Departamento de F\'{\i}sica, Universidade
Federal da Para\'{\i}ba,\\
Caixa Postal 5008, 58059-970 Jo\~{a}o Pessoa, Pb, Brazil.}

\begin{abstract}
We develop a stochastic approach to study gravitational waves
produced during the inflationary epoch under the presence of a
decaying cosmological parameter, on a 5D geometrical background
which is Riemann flat. We obtain that the squared tensor metric
fluctuations
depend strongly on the cosmological parameter 
$\Lambda (t)$ and we finally illustrate the formalism with an
example of a decaying $\Lambda(t)$.
\end{abstract}

\maketitle

\section{Introduction}

Currently, embedding theorems are of rather interest in much due to their property of
constraining the ways on which general 
relativity can be welded to higher dimensional scenarios, which may in principle embody
internal symmetry groups like those that 
appear in particle physics. An interesting example of these are the
Campbell-Magaard theorem and its known extensions that provide a 
kind of ladder between manifolds whose dimensionality differs just by one \cite{campbell}.
One of the most valuable physical 
implications of these theorems are that they guarantee that any physical source of
4D matter can be geometrically modeled from a 
Ricci-flat, Einstein or scalar field sourced higher dimensional manifolds.
These powerful theorems make relevant the study of 
different cosmological topics on these settings. In particular to investigate in
a stochastic approach the behavior of 
gravitational waves under the presence of a decaying cosmological parameter
on a 5D Ricci-flat space-time can lead physics of 
vanguard in the study of cosmology of the early universe.\\

In the stochastic approach to inflation quantum to classical
transition dynamics of the scalar field (inflaton) is effectively
described by a classical noise, which has  quantum origin
\cite{habib,mijic,BCMS}. This transition effect can be also
studied by employing scalar metric fluctuations \cite{anabitarte}.
In this paper we investigate this transition dynamics in the case
of linearized tensor perturbations \cite{bar}, which as we know
describe gravitational waves during inflation. Within the
inflationary theory the  prediction of the existence of a
background of gravitational waves arises naturally \cite{1}. These
tensor perturbations escape out of the horizon during inflation,
remaining this way completely conserved to form a relic of
background gravitational waves, which carries out information of
the very early universe \cite{2,3,4}. This paper seeks to be a
continuation of a recently introduced formalism where we have
studied gravitational waves from a 5D vacuum state \cite{edgar},
considering an accelerated expansion of the universe governed by a
decreasing cosmological parameter $\Lambda(t)$ during inflation
\cite{plb2007}. Five dimensions are of particular interest, since
this represents the simplest extension of spacetime and is widely
regarded as the low-energy limit of even higher-dimensional
theories with relevance to particle physics, such as a $10D$
supersymmetry, $11D$ supergravity and higher-$D$ versions of
string theory. However, our approach is inspired in the Induced
Matter theory (STM) \cite{STM}, where 4D sources appear as induced
by one extended extra dimension, meaning, by extended, that the
fifth dimension is considered noncompact. In the present approach
the components of the tensor metric fluctuations are
coarse-grained with an increasing number of degrees of freedom. In
consequence the dynamics of the components of the coarse-grained
tensor field are described by a set of second order stochastic
equations, which can be rewritten as two sets of first order.\\

The paper is organized as follows: in Sect. II we introduce the
formalism for tensor metric fluctuations on a 5D Riemann flat
metric. In Sect. III we describe an effective 4D dynamics for
these fluctuations when we take a foliation $\psi = \psi_0$ on the
fifth coordinate. Furthermore, we describe the dynamics of the
coarse-grained field, which describes the tensor fluctuations on
super-Hubble scales, which are the relevant scales for cosmology
during the inflationary evolution of the universe. In particular,
we study the evolution of the squared $^{(L)} h$-field
fluctuations and its spectrum. Finally, in Sect. IV we develop
some conclusions and final comments.\\

\section{GW in a 5D vacuum: preliminary formalism}

In this section we shall establish the basic formalism with the
necessary ingredients in order to construct a stochastic formalism
able to describe GW from a 5D vacuum in the presence of a time
dependent cosmological parameter $\Lambda (t)$. With this aim we
start considering the background line element \cite{M1}
\begin{equation}\label{eq1}
dS_{b}^{2}=\psi^{2}\frac{\Lambda (t)}{3}dt^{2}-\psi^{2}e^{2\int
\sqrt{\Lambda/3}\,dt}dr^2 - d\psi^{2},
\end{equation}
where $dr^{2}=\delta _{ij}dx^{i}dx^{j}$,
being $\lbrace x^{i} \rbrace =\lbrace x,y,z\rbrace$
the local cartesian coordinates. Here $t$ is the cosmic time and $\psi$ is the fifth coordinate which is
space-like. Adopting a natural unit system (where $\hbar =c=1$), the fifth coordinate $\psi$ has spatial
units whereas the cosmological parameter $\Lambda (t)$ has units of $(length)^{-2}$.
The background metric in (\ref{eq1})
is Riemann-flat, $R^{A}\,_{BCD}=0$ and thereby it describes a 5D geometrical  vacuum.\\

The second order 5D action for the tensor perturbations in our
case is
\begin{equation}\label{neq1}
^{(5)} {\cal S}=  \int d^{4}x \, d\psi  \sqrt{\left|\frac{^{(5)} g}{^{(5)} g_0}\right|}
\left[ \frac{^{(5)} {\cal R}}{16\pi G} + \frac{M^2_p }{2} g^{AB}
Q^{ij}_{\, ;A} Q_{ij;B}\right],
\end{equation}
being $Q_{ij}(t,\vec{r},\psi)$ the transverse traceless tensor
denoting the tensor fluctuations with respect to the background
metric $g_{AB}$, and therefore the expressions
$tr(Q_{ij})=Q^{i}\,_{i}=0$ and $Q^{ij}\,_{;i}=0$ are valid. In
addition the comma $(;)$ is denoting covariant derivative and
$^{(5)}g_{0}= ^{(5)}g [\psi =\psi _{0},\Lambda _{0}= \Lambda
(t=t_{0})]$ is a dimensionalization constant being $\psi _0$ and
$t_0$ some constants to be specified. From the action (\ref{neq1})
we derive the evolution equation for the tensor fluctuations
$Q_{ij}$
\begin{equation}\label{eq4}
\ddot{Q}_{ij} + \left[3\sqrt{\frac{\Lambda}{3}}-
\frac{1}{2}\frac{\dot{\Lambda}}{\Lambda}\right]\dot{Q}_{ij}-\frac{\Lambda}{3} \  e^{-2\int \sqrt{\Lambda/3}\,dt}
\nabla^2_r
Q_{ij}-\frac{\Lambda}{3}\left[4\psi Q_{ij,\psi}+\psi^{2}Q_{ij,\psi,\psi}\right]=0,
\end{equation}
where the dot denotes derivative with respect to the cosmic time $t$.
Quantization of $Q_{ij}$ is achieved by demanding the commutation relation
\begin{equation}\label{com}
\left[Q_{ij}(t,\vec{r},\psi),\frac{\partial
\,^{(5)}L^{(GW)}}{\partial Q_{ls,t}}(t,\vec{r'},\psi)\right]=i
\delta^l_i \ \delta^s_j \
 g^{tt} M^2_p \sqrt{\left|\frac{^{(5)}g}{^{(5)}g_0}\right|}
\left(\frac{\psi_0}{\psi}\right)^3
e^{-\int\left[3\sqrt{\frac{\Lambda}{3}} -
\frac{\dot\Lambda}{2\Lambda}\right] dt} \,\,
\delta^{(3)}(\vec{r}-\vec{r'}),
\end{equation}
and expressing the quantum operators $Q_{ij}$ as a
Fourier expansion of the form
\begin{equation}\label{eq5}
Q^{i}\,_{j}(t,\vec{r},\psi)=\frac{1}{(2\pi)^{3/2}}\int
d^{3}k_{r}\,\sum _{\alpha}\,^{(\alpha)} e^{i}\,_{j}\left[
e^{i\vec{k}_{r}\cdot\vec{r}}\zeta^{(\alpha)} _{k_{r}}(t,\psi)+
e^{-i\vec{k}_{r}\cdot\vec{r}}\left(\zeta^{(\alpha)}_{k_{r}}(t,\psi)\right)^{*}\right],
\end{equation}
with $\alpha $ counting the number of polarization degrees of
freedom and the asterisk $(*)$ denoting complex conjugate. The
polarization tensor $^{(\alpha)}e_{ij}$ obeys
\begin{equation}\label{eq7}
^{(\alpha)}e_{ij}=\,^{(\alpha)}e_{ji},\quad ^{(\alpha)}
e_{ii}=0,\quad k^{i}\,^{(\alpha)}e_{ij}=0,\quad ^{(\alpha)}e_{ij}(-\vec{k}_r)=\,^{(\alpha)}e_{ij}^{*}(\vec{k}_r).
\end{equation}
On the other hand, by introducing the quantities
\begin{equation}\label{neq2}
\chi _{ij}(t,\vec{r},\psi)=e^{\frac{1}{2}\int 
\left[3\sqrt{\frac{\Lambda}{3}}-\frac{\dot{\Lambda}}{2\Lambda}\right]dt}\,Q_{ij}(t,\vec{r},\psi),
\end{equation}
the evolution equation (\ref{eq4}) becomes
\begin{equation}\label{neq3}
\ddot{\chi}^{i}\,_{j}-\frac{\Lambda}{3}\,e^{-2\int \sqrt{\Lambda/3}dt}\nabla _{r}^{2}\chi 
^{i}\,_{j}+\left[\frac{1}{4}\frac{\ddot{\Lambda}}{\Lambda}-\frac{3}{4}\Lambda-\frac{5}{16}
\frac{\dot{\Lambda}^2}{\Lambda^2}\right]\chi ^{i}\,_{j}-\frac{\Lambda}{3}\left[4\psi\frac{\partial}{\partial 
\psi}+\psi^{2}\frac{\partial^{2} }{\partial\psi ^2}\right]\chi ^{i}\,_{j}=0.
\end{equation}
Inserting the expansion (\ref{eq5}) in (\ref{neq2}) we obtain the Fourier
expansion for the quantum operators $\chi ^{i}\,_{j}$
\begin{equation}\label{neq4}
\chi ^{i}\,_{j}(t,\vec{r},\psi)=\frac{1}{(2\pi)^{3/2}}\int
d^{3}k_{r}\,\sum _{\alpha}\,^{(\alpha)} e^{i}\,_{j}\left[
e^{i\vec{k}_{r}\cdot\vec{r}}\, \xi^{(\alpha)}_{k_{r}}(t,\psi)+
e^{-i\vec{k}_{r}\cdot\vec{r}}\,
\left(\xi^{(\alpha)}_{k_{r}}(t,\psi)\right)^{*}\right],
\end{equation}
where we have introduced the re-defined modes
\begin{equation}\label{neq5}
\xi^{(\alpha)} _{k_{r}}(t,\psi)=e^{\frac{1}{2}\int
\left[3\sqrt{\frac{\Lambda}{3}}-\frac{\dot{\Lambda}}{2
\Lambda}\right]dt}\zeta^{(\alpha)}_{k_{r}}(t,\psi).
\end{equation}
The dynamical equation for the re-defined modes then reads
\begin{equation}\label{neq6}
\ddot{\xi}^{(\alpha)}_{k_{r}}+\left[\frac{\Lambda}{3}\,e^{-2\int\sqrt{\Lambda}{3}dt}k_{r}^{2}+
\frac{1}{4}\frac{\ddot{\Lambda}}{ \Lambda}-\frac{3}{4}\Lambda -
\frac{5}{16}\frac{\dot{\Lambda}^{2}}{\Lambda ^2}\right]\xi^{(\alpha)} 
_{k_{r}}-\frac{\Lambda}{3}\left[4\psi \frac{\partial}{\partial
\psi}
+\psi^{2}\frac{\partial ^{2}}{\partial 
\psi^2}\right]\xi^{(\alpha)} _{k_{r}}=0.
\end{equation}

The equation (\ref{neq6}) yields
\begin{eqnarray}
&& \ddot{\xi}_{k_r}+\left[\frac{\Lambda}{3}\,e^{-2\int \sqrt{\Lambda/3}\,dt} k_{r}^{2}-
\frac{1}{4}\sqrt{\frac{3}{\Lambda}}\dot{\Lambda}+\frac{1}{4}
\frac{\ddot{\Lambda}}{ \Lambda}-\frac{5}{16}\frac{\dot{\Lambda}^2}{\Lambda^2}
+\frac{3}{4}\sqrt{\frac{\Lambda}{3}}\frac{\dot{\Lambda}}{\Lambda}+\left( \frac{m^2}{3}-
\frac{3}{4}\right)\Lambda \right]\xi _{k_r}=0, \label{eq12}\\
&& \frac{d^2 L_m(z)}{dz^2} + \left[m^2 - \frac{9}{4}\right] L_m(z) =0,
\label{eq12'}
\end{eqnarray}
where  we have used the transformation $\Theta
_{m}(z)=e^{-(3/2)z}L_{m}(z)$ with $z=ln(\psi/\psi _0)$ and the
parameter $m^{2}$ is a separation constant which is related with
the squared of the KK mass measured by a class of 5D observers.
This way, given a cosmological parameter $\Lambda (t)$, the
temporal evolution of the tensor modes $\xi _{k_r}(t)$ in 5D is
determined by solutions of (\ref{eq12}). Once solutions for $\xi
_{k_r}(t)$ and $L_{m}(z)$ are obtained, they should satisfy the
algebra (\ref{com}). This can be achieved if the solutions satisfy
respectively the condition
\begin{equation}\label{eq13}
{\Large\int} d^3k_r\,\,\left<0\left|\left[\xi^{(\alpha)}
_{k_{r}}\left(\dot{\xi}^{(\alpha)}_{k_{r}}\right)^{*}-
\dot{\xi}^{(\alpha)}_{k_{r}}\left(\xi
_{k_{r}}^{(\alpha)}\right)^{*}\right]\right|0\right>=i,
\end{equation}
which are usually named the normalization conditions.

On the other hand, note that equation (\ref{eq12'}) is exactly the
same as the one obtained in \cite{edgar}. Therefore about the
behavior of the modes with respect the fifth coordinate we can say
that for $m>3/2$ the KK-modes are coherent on the ultraviolet
sector (UV), described by the modes
\begin{equation}\label{eq14}
k^2_r > k^2_0(t) = \left\{\frac{3}{2\Lambda} \frac{d}{dt}\left[ 3
\sqrt{\frac{\Lambda}{3}} - \frac{\dot\Lambda}{2\Lambda}\right]
-\frac{3}{4\Lambda}
\left(3\sqrt{\frac{\Lambda}{3}}-\frac{\dot\Lambda}{2\Lambda}\right)^2
- m^2\right\} \  e^{2\int \sqrt{\frac{\Lambda}{3}} dt} >0.
\end{equation}
Notice that in the case of a constant cosmological parameter
$\Lambda = \Lambda_0$, we obtain that
\begin{displaymath}
k^2_0(t) = \left( \frac{9}{4} - m^2 \right) \,\,e^{2\int
\sqrt{\Lambda/3} dt},
\end{displaymath}
and the line element (\ref{eq1}) give us the Ponce de Le\'on
metric\cite{pdl}. Notice that for $m > 3/2$ the solutions of
(\ref{eq12}) are always stable. However, for $m<3/2$ those modes
are unstable and diverge at infinity when $k_r < k_0$. The modes
with $m > 3/2$ comply with the conditions (\ref{eq13}), so that
they are normalizable.

The coarse-grained tensor field which describes a stochastic dynamics on super-Hubble scales is defined by
\begin{equation}\label{cg1}
\chi ^{i}\,_{j}(t,\vec{r},\psi)=\frac{1}{(2\pi)^{3/2}}\int
d^{3}k_{r}\,\theta(\epsilon k_0 -k_r)\,\sum _{\alpha}\,^{(\alpha)}
e^{i}\,_{j}\left[ e^{i\vec{k}_{r}\cdot\vec{r}}\,
\xi^{(\alpha)}_{k_{r}}(t,\psi)+ e^{-i\vec{k}_{r}\cdot\vec{r}}\,
\left(\xi^{(\alpha)}_{k_{r}}(t,\psi)\right)^{*}\right],
\end{equation}
This field $^{(L)}\chi ^{i}\,_{j}(t,\vec{r},\psi)$ contains all
the modes in the IR-sector $k_{r}/k_{0} <\epsilon \simeq 10
^{-3}$. This means that $^{(L)}\chi ^{i}\,_{j}$ only considers
modes with wavelengths larger than $10^{3}$ times the size of the
horizon during inflation.

\section{Effective 4D dynamics}

As in a recently introduced work \cite{plb2007} we shall assume that the 5D space-time
can be foliated by a family of 
hypersurfaces where a generic hypersurface is determined by taking
$\psi=\psi_0$. An extension to dynamical foliations was recently
studied in\cite{e-print}. Thus, the line element (\ref{eq1})
generates an effective 4D background metric
$\left.dS^2\right|_{eff} = ds^2$, where
\begin{equation}\label{61}
ds^2 = \psi^2_0 \frac{\Lambda(t)}{3} dt^2 - \psi^2_0 e^{2 \int
\sqrt{{\Lambda\over 3}}dt} d\vec{r}^2.
\end{equation}
It is important to notice that when the cosmological parameter is
constant, $\Lambda=\Lambda_0$, the effective 4D metric describes a
de Sitter expansion with an energy density $\rho_v
={\Lambda_0\over 8\pi G}=-{\rm p}_v$, where $\rho_v$ and ${\rm
p}_v$ are respectively the energy density and the pressure on a
vacuum dominated expansion. In particular, when we use the
foliation $\psi_0=\sqrt{3/\Lambda_0}=H^{-1}_0$ (in this case $H_0$
is the constant Hubble parameter), we obtain a comoving reference
system with tetra-velocities $u^{\alpha} = (1,0,0,0)$ and the
universe can be described by a Friedmann-Robertson-Walker (FRW)
metric $ds^2 = dt^2 - a^2(t) d\vec{r}^2$ [$a^2 = H^{-2}_0 e^{2
H_0\, t}$] with an exponential (vacuum dominated) expansion. In
general [i.e., when $\Lambda=\Lambda(t)$], the effective 4D metric
(\ref{61}) is not the usual FRW one, and the tetra-velocities
$u^t$ and $u^r$ are related by the expression
\begin{displaymath}
\left(u^t\right)^2 = \frac{3}{\Lambda} \left[\psi^{-2}_0 + \left(
u^r\right)^2 e^{-2\int \sqrt{\Lambda/3} dt} \right],
\end{displaymath}
where $\left( u^r\right)^2 = \left( u^x\right)^2 + \left(
u^y\right)^2 + \left( u^z\right)^2$.

On the other hand, the dynamics of the 4D tensor-fluctuations will
be given in terms of the tensor components
$h_{ij}(t,\vec{r})\equiv Q_{ij}(t,\vec{r},\psi=\psi_0)$. The
effective 4D action ($\alpha$, $\beta$ run from $0$ to $3$) can be
written as
\begin{equation}\label{act}
^{(4)} {\cal S} = -\int d^4 x \sqrt{\left|\frac{^{(4)} g}{^{(4)}
g_0}\right|} \left. \left[ \frac{^{(4)} {\cal R}}{16\pi G} +
\frac{M^2_p }{2} g^{\alpha\beta}  Q^{ij}_{\, ;\alpha}
Q_{ij;\beta}\right] \right|_{\psi=\psi_0},
\end{equation}
where $^{(4)} {\cal R} = 12/\psi^2_0$ is the effective 4D Ricci
scalar evaluated on the metric (\ref{61}). In other words, the 4D
scalar curvature is geometrically induced by the foliation on the
fifth coordinate: $\psi = \psi_0$.\\

The effective 4D linearized equation of motion for the 4D
tensor-fluctuations is
\begin{equation}\label{eq18}
\left.\ddot{h}^i_j + \left[3\sqrt{\frac{\Lambda}{3}} -
\frac{\dot\Lambda}{2\Lambda} \right] \dot{h}^i_j -
\frac{\Lambda}{3} e^{-2\int\sqrt{\frac{\Lambda}{3}} dt} \nabla^2_r
h^i_j + \frac{\Lambda}{3}\left[4\psi\frac{\partial}{\partial
\psi}+\psi^{2}\frac{\partial^{2} }{\partial\psi ^2}\right]
h^i_j\right|_{\psi=\psi_0} =0,
\end{equation}
which, after make the transformation $h^i_j(t,\vec r) = e^{-1/2
\int \left[ 3\left({\Lambda\over 3}\right)^{1/2}-{\dot\Lambda\over
2\Lambda}\right] dt} \chi^i_j(t,\vec r)$, give us the equation of
motion for the redefined 4D tensor-fluctuations
\begin{equation}\label{eq19}
\left.\ddot{\chi}^i_j - \frac{\Lambda}{3}
e^{-2\int\sqrt{\frac{\Lambda}{3}} dt} \nabla^2_r \chi^i_j - \left[
\frac{1}{4} \sqrt{\frac{3}{\Lambda}} \dot\Lambda + \frac{1}{4}
\frac{\ddot\Lambda}{\Lambda} - \frac{5}{16}
\frac{\dot\Lambda^2}{\Lambda^2} + \frac{3}{4}
\sqrt{\frac{\Lambda}{3}} \frac{\dot\Lambda}{\Lambda} + \left[
\frac{1}{3}\left(4\psi\frac{\partial}{\partial
\psi}+\psi^{2}\frac{\partial^{2} }{\partial\psi ^2}\right) -
\frac{3}{4} \right]  \Lambda \right] \chi^i_j\right|_{\psi=\psi_0}
=0.
\end{equation}
Therefore, for a given $\Lambda (t)$ we can obtain in principle an effective
4D dynamics for the tensor fluctuations of the 
metric. Now, instead of following the standard procedure to
investigate the 4D effective dynamics of this
tensor modes, let us to 
adopt a stochastic approach.

\subsection{Coarse grained field in 4D}

In order to describe the tensor-fluctuations on cosmological
scales, we shall introduce the coarse-grained tensor field
$^{(L)}\chi ^{i}\,_{j}\left(t,\vec{r},\psi=\psi_0\right)$ on the
effective 4D metric (\ref{61}). This field is represented as a
Fourier expansion on the modes whose wavelengths are bigger than
the Hubble radius
\begin{eqnarray}
^{(L)}\chi^{i}\,_{j}\left(t,\vec{r},\psi_0\right)& = &
\frac{1}{(2\pi)^{3/2}}\int d^{3}k_{r} \,\theta (\epsilon
k_{0}-k_{r})\sum _{\alpha}^{}\,^{(\alpha)}e^{i}\,_{j} \nonumber
\\
& \times & \left[e^{i\vec{k}_{r}\cdot\vec{r}}
\,\xi^{(\alpha)}_{k_{r}}(t,\psi_0) +
e^{-i\vec{k}_{r}\cdot\vec{r}}\left(\xi^{(\alpha)}_{k_{r}}(t,\psi_0)\right)^{*}
\right],
\end{eqnarray}
where, because on the effective 4D hypersurface $\psi$ is a
constant $\psi_0$, we shall consider that $m$ is a free parameter,
such that
\begin{equation}
\xi^{(\alpha)}_{k_{r}}(t,\psi_0)= a^{(\alpha)}_{k_r}
\xi_{k_{r}}(t), \qquad
\left[a_{k_{r}}^{(\alpha)},a_{k'_{r}}^{(\alpha ')\,\,\dagger}\right]=g^{\alpha\alpha '}\delta 
^{(3)}(\vec{k}_{r}-\vec{k}'_{r}),
\end{equation}
and
\begin{equation}\label{op1}
k_{0}(t)=\left[\frac{5}{16}\frac{\dot{\Lambda}^2}{\Lambda 
^2}-\frac{1}{4}\frac{\ddot{\Lambda}}{\Lambda}+\left(\frac{3}{4}-\frac{m^{2}}{3}\right)\Lambda\right]^{1/2}
\sqrt{\frac{3}{\Lambda}}\,\,e^{\int\sqrt{\frac{ \Lambda}{3}}dt}.
\end{equation}
The equation of motion for the field
$^{(L)}\chi^{i}\,_{j}\left(t,\vec{r},\psi_0\right)$ is
\begin{equation}\label{cg31}
^{(L)}\ddot{\chi}^{i}\,_{j} - \frac{\Lambda}{3}\,k_{0}^{2}\,e^{-2\int\sqrt{\frac{\Lambda}{3}}dt}\,^{(L)}\chi 
^{i}\,_{j}=\epsilon\left[\ddot{k}_{0}\eta ^{i}\,_{j}(t,\vec{r},\psi_0)+\dot{k}_{0}\kappa 
^{i}\,_{j}(t,\vec{r},\psi_0)+2\dot{k}_{0}\gamma
^{i}\,_{j}(t,\vec{r},\psi_0)\right],
\end{equation}
where the stochastic tensor operators $\eta ^{i}\,_{j}$, $\kappa
^{i}\,_{j}$ and $\gamma ^{i}\,_{j}$, on the effective 4D metric
(\ref{61}), are given, respectively, by
\begin{eqnarray}\label{cg41}
\eta ^{i}\,_{j}(t,\vec{r},\psi_0)&=&\frac{1}{(2\pi)^{3/2}}\int
d^{3}k_{r}\,
\delta (\epsilon k_{0}-k_{r})\sum _{\alpha}\,^{(\alpha)}e^{i}\,_{j} \nonumber \\
&\times & \left[e^{i\vec{k}_{r}\cdot\vec{r}}\,\xi^{(\alpha)}_{k_{r}}(t,\psi_0) + 
e^{-i\vec{k}_{r}\cdot\vec{r}}\left(\xi^{(\alpha)}_{k_{r}}(t,\psi_0)\right)^{*} \right],\\
\label{cg51} \kappa
^{i}\,_{j}(t,\vec{r},\psi_0)&=&\frac{1}{(2\pi)^{3/2}}\int
d^{3}k_{r}\,
\dot{\delta}(\epsilon k_{0}-k_{r})\sum _{\alpha}\,^{(\alpha)}e^{i}\,_{j} \nonumber \\
& \times  & \left[e^{i\vec{k}_{r}\cdot\vec{r}}\,\xi^{(\alpha)}_{k_{r}}(t,\psi_0) + 
e^{-i\vec{k}_{r}\cdot\vec{r}}\left(\xi^{(\alpha)}_{k_{r}}(t,\psi_0)\right)^{*} \right],\\
\label{cg61} \gamma
^{i}\,_{j}(t,\vec{r},\psi_0)&=&\frac{1}{(2\pi)^{3/2}}\int
d^{3}k_{r}\,\delta (\epsilon k_{0}-k_{r})\sum
_{\alpha}\,^{(\alpha)}e^{i}\,_{j} \nonumber
\\
& \times & \left[e^{i\vec{k}_{r}\cdot\vec{r}}\,
\dot{\xi}^{(\alpha)}_{k_{r}}(t,\psi_0) +
e^{-i\vec{k}_{r}\cdot\vec{r}}\left(\dot{\xi}^{(\alpha)}_{k_{r}}(t,\psi_0)\right)^{*}
\right].
\end{eqnarray}

By using differential properties of the former stochastic
operators, the equation (\ref{cg31}) can be written as
\begin{equation}\label{cg7}
^{(L)}\ddot{\chi}^{i}\,_{j}-\frac{\Lambda}{3}k_{0}^{2}\,e^{-2\int\sqrt{\frac{\Lambda}{3}}\,dt}\,^{(L)}\chi 
^{i}\,_{j}=\epsilon\left[\frac{\partial}{\partial
t}\left(\dot{k}_{0}\eta ^{i}\,_{j}(t,\vec{r},\psi_0)\right)
+\dot{k}_{0}\gamma ^{i}\,_{j}(t,\vec{r},\psi_0)\right].
\end{equation}
This is a second-order stochastic equation that can be written as
a first-order system in the form
\begin{eqnarray}\label{cg8}
\dot{u}^{i}\,_{j}&=& \frac{\Lambda}{3}k_{0}^{2}\,e^{-2\int\sqrt{\frac{\Lambda}{3}}\,dt}\,^{(L)}\chi ^{i}\,_{j}
+ \epsilon \dot{k}_{0}\gamma ^{i}\,_{j},\\
\label{cg9} ^{(L)}\dot{\chi} ^{i}\,_{j}&=&
u^{i}\,_{j}+\epsilon\dot{k}_{0} \eta^{i}\,_{j} ,
\end{eqnarray}
where we have introduced the auxiliary field
$u^{i}\,_{j}\equiv\,^{(L)}\dot{\chi} ^{i}\,_{j} -\epsilon
\dot{k}_{0}\eta ^{i}\,_{j}$. Now, in order to minimize the role of
the stochastic noise $\gamma ^{i}\,_{j}$, we impose the condition
$\dot{k}_{0}^{2}\left<\gamma^{2}\right> \ll
\ddot{k}_{0}^{2}\left<\eta ^{2}\right>$, where we have defined the
quantities $\left<\gamma ^{2}\right>=\,\left<0|\gamma
^{i}\,_{j}\gamma _{i}\,^{j}|0\right>$ and $\left<\eta
^{2}\right>=\,\left<0|\eta ^{i}\,_{j}\eta _{i}\,^{j}|0\right>$.
This condition can be expressed in terms of the modes as
\begin{equation}\label{cg10}
\left.\frac{\dot{\xi}^{(\alpha)}_{k_{r}}\left(\dot{\xi}^{(\alpha)}_{k_{r}}\right)^{*}}{\xi^{(\alpha)}
_{k_{r}}\left(\xi^{(\alpha)}
_{k_{r}}\right)^{*}}\right|_{\psi=\psi_0} \ll \left(\frac{\ddot
k_0}{\dot k_0}\right)^2,
\end{equation}
which only is valid on super-Hubble scales. Under this
consideration the system (\ref{cg8})-(\ref{cg9}) becomes
\begin{eqnarray}\label{cg11}
\dot{u}^{i}\,_{j}&=& \frac{\Lambda}{3}k_{0}^{2}\,e^{-2\int\sqrt{\frac{\Lambda}{3}}\,dt}\,^{(L)}\chi ^{i}\,_{j} ,\\
\label{cg9n} ^{(L)}\dot{\chi} ^{i}\,_{j}&=&
u^{i}\,_{j}+\epsilon\dot{k}_{0} \eta^{i}\,_{j}.
\end{eqnarray}
This new system can be seen as two firsth-order Langevin equations
with a tensor noise $\eta ^{i}\,_{j}$ which is Gaussian and white
in nature. Hence, it satisfies
\begin{eqnarray}\label{cg12}
\left<\eta\right>&=& \left<g^j\,_{i} \eta^i\,_{j}\right>=0,\\
\label{cg13} \left.\left<\eta ^{2}\right> \right|_{\psi=\psi_0}
&=& \left.\left< \eta^i\,_{j}
\eta^j\,_{i}\right>\right|_{\psi=\psi_0} =\frac{3\epsilon
k_{0}^{2}}{\pi ^{2}\dot{k}_{0}} \,\,\xi _{\epsilon k_{0}}(t)
\,\xi^{*}_{\epsilon k_{0}}(t)
 \,\delta (t-t') .
\end{eqnarray}
The corresponding Fokker-Planck equation that describes the
dynamics of the transition probability ${\cal
P}^i\,_{j}[\,^{(L)}\chi_{0}\,
^{i}\,_{j},u_{0}\,^{i}\,_{j}|\,^{(L)}\chi ^{i}\,_{j},u^{i}\,_{j}]$
from a configuration $(\,^{(L)}\chi
_{0}\,^{i}\,_{j},u_{0}\,^{i}\,_{j})$ to $(\,^{(L)}\chi
^{i}\,_{j},u^{i}\,_{j})$ is then
\begin{equation}\label{cg14}
\frac{\partial {\cal P}^i\,_{j}}{\partial
t}=-u^{i}\,_{j}\frac{\partial {\cal P}^i\,_{j}}{
\partial \,^{(L)\chi ^{i}\,_{j}}}-\mu ^{2}(t)\,^{(L)}\chi ^{i}\,_{j}\frac{\partial {\cal
P}^i\,_{j}}{\partial u^{i}\,_{j}}+\frac{1}{6}
D_{\eta\eta}\frac{\partial ^{2}{\cal P}^i\,_{j}}{\partial
(\,^{(L)}\chi ^{i}\,_{j})^{2}}\, ,
\end{equation}
where $\mu ^{2}(t)=(\Lambda/3)k_{0}^{2}\exp
[-2\int\sqrt{\Lambda/3}\,dt]$ and the only nonzero component of
the diffusion tensor is
$D_{\eta\eta}(t)=[(\epsilon\dot{k}_{0}^{2})/2]\,\int dt
\left<\eta^{2}\right>$. Note that we have considered that
$D_{\eta\eta}=3 D_{\eta^i\,_{j}\,\eta^j\,_{i}}$, due to the 3D
space $r(x,y,z)$ is isotropic. This diffusion coefficient is
related to the field $^{(L)}\chi $ due to the stochastic action of
the effective noise $\eta$ (related to $\eta^i\,_{j})$. The
equation of motion for $\left<\,^{(L)}\chi ^{2}\right> \equiv
\left<0|\,^{(L)}\chi ^{i}\,_{j}\,^{(L)}\chi _{i}\,^{j}|0\right>
=\int d\,^{(L)}\chi \, du\,^{(L)}\chi\,u{\cal P}
[\,^{(L)}\chi,u]$, takes the form
\begin{equation}\label{cg16}
\frac{d}{dt}\left<\,^{(L)}\chi \,^{2}\right>=\frac{1}{2}
D_{\eta\eta}(t),
\end{equation}
where we have considered $g^j\,_{i} {\cal P}^i\,_{j}={\cal P}$,
$\chi = g^j\,_{i} \chi^i\,_{j}$ and $u = g^j\,_{i} u^i\,_{j}$.
Therefore, the stochastic dynamics of $^{(L)}\chi$ is completely
determined and consequently the corresponding evolution of
$\left<^{(L)}h^2 \right>$ is given by the solution of
\begin{equation}
\frac{d}{dt} \left<^{(L)} h^2\right> =
-\left[3\sqrt{\frac{\Lambda}{3}} -
\frac{\dot\Lambda}{2\Lambda}\right] \left<^{(L)} h^2\right>
+\frac{1}{2} e^{-\int dt \left[3\sqrt{\frac{\Lambda}{3}} -
\frac{\dot\Lambda}{2\Lambda}\right]} D_{\eta\eta}(t).
\end{equation}
The general solution of this equation is
\begin{equation}\label{va1}
\left<\,^{(L)}h^{2}\right>=\frac{1}{2}\,e^{-\int dt
\left[3\sqrt{\frac{\Lambda}{3}} -
\frac{\dot\Lambda}{2\Lambda}\right]}\int D_{\eta\eta}(t)\, dt
\end{equation}
where $D_{\eta\eta}=[3\epsilon ^3\dot{k}_{0}k_{0}^{2}/(2\pi^{2})]
\xi _{\epsilon k_{0}}(t) \xi^{*} _{\epsilon k_{0}}(t)$. Therefore,
for a given cosmological parameter $\Lambda (t)$, the stochastic
squared fluctuations of $h$ on super Hubble scales, $^{(L)}h$, is
determined by (\ref{va1}).

\subsection{An example}

Now let us to illustrate the previous formalism by considering a
decaying cosmological parameter on the metric (\ref{61}): $\Lambda
(t)=3p^{2}/t^{2}$ (with $p>0$), which clearly satisfies
$\dot{\Lambda}<0$. In this
case the dynamical field equation for 
the modes $\xi_{k_{r}}(t)$ reads
\begin{equation}\label{va2}
\ddot{\xi}_{k_{r}}+\left\lbrace 
k_{r}^{2}p^{2}t_{0}^{2p}t^{-2(p+1)}-\left[\left(m^{2}-\frac{9}{4}\right)p^{2}-\frac{9}{4}p+\frac{1}{4}
\right]t^{-2}\right\rbrace \xi _{k_{r}}=0,
\end{equation}
where $M^{2}(t)=[(m^{2}-9/4)p^{2}-(9/4)p+1/4]t^{-2}$ can be interpreted as an effective squared term
of mass. The permitted values for $m$ should be $9/4<m^{2}\leq (9/4)[1+9/4]$, for which the next relation is valid
\begin{equation}\label{va3}
0 < p\leq \frac{9+\sqrt{117-16m^{2}}}{2(4m^{2}-9)}.
\end{equation}
However, as it was shown in \cite{plb2007}, even when the general
solution of (\ref{va2}) is not normalizable, there exist some
particular normalizable solutions. One class of these solutions
is obtained by considering 
$M^2(t) \geq 0$. In this case the expression (\ref{va2}) reduces
to
\begin{equation}\label{va4}
\ddot{\xi}_{k_{r}}+\left[k_{r}^{2}p^{2}t_{0}^{2p}t^{-2(p+1)}-M^2(t)\right]\xi_{k_{r}}=0,
\end{equation}
whose normalized solution (using the Bunch-Davies vacuum\cite{BD}
), is given by
\begin{equation}\label{va5}
\xi _{k_{r}}(t)=\frac{i}{2}\sqrt{\frac{1}{\pi p}}\sqrt{t}\,{\cal 
H}_{\nu}^{(2)}\left[k_{r}\left(\frac{t_0}{t}\right)^{p}\right]
\end{equation}
where ${\cal H}_{\nu}^{(2)}$ is the second kind Hankel function
and $\nu^2 =m^2-\left[9p(p+1)-2\right]/(4p^2)$. Thus considering
that in this particular case $k_{0}(t)=\sqrt{\alpha
/p}\,(t/t_0)^{p}$ and using the asymptotic expansion ${\cal
H}_{\nu}^{(2)}[x]\simeq (-i/\pi)\Gamma (\nu)[x/2]^{-\nu}$, the
diffusion coefficient $D_{\eta\eta}$ has the form
\begin{equation}\label{va6}
D_{\eta\eta}(t)=\frac{3\epsilon ^{3-2\nu}}{\pi ^{5}p^{3/2-\nu}}\,2^{-(3-2\nu)}\,\alpha ^{3/2-\nu}\,\Gamma 
^{2}(\nu)\,\left(\frac{t}{t _0}\right)^{3p}\, ,
\end{equation}
where $\alpha=M^2(t) t^2/p^2\geq 0$ is a real constant. Hence
equation (\ref{va1}) gives
\begin{equation}
\left<\,^{(L)}h ^{2}\right>=\frac{3\epsilon ^{3-2\nu}}{\pi ^{5}(3p+1)p^{3/2-\nu}}\,\Gamma ^{2}(\nu)
\,2^{-4+2\nu}\,\alpha 
^{3/2-\nu}\,t_{0}\,\left[1-\left(\frac{t_0}{t}\right)^{1+3p}\right],
\end{equation}
which, for $p>0$ and $t>t_0$ is always a positive quantity. Note
that when $\alpha =0$ automatically $\left<\,^{(L)}h
^{2}\right>=0$. On the other hand, for a scale invariant spectrum
(for which $\nu =3/2$) we have
\begin{equation}\label{vva1}
\left<\,^{(L)}h ^{2}\right>=\frac{3\Gamma ^{2}(\nu)}{2\pi ^{5}(3p+1)}\,t_{0}
\,\left[1-\left(\frac{t_0}{t}\right)^{1+3p}\right],
\end{equation}
which is independent of the value of $\alpha$. In this case we
have that $9p^2-4\alpha p -1=0$, so that $p = {2\alpha\over 9}
\left[1+ \sqrt{1+ {9\over 4\alpha^2}}\right]$, for $\alpha >0$.

In general, the power spectrum of $\left<^{(L)} h^2 \right> $ has
the form
\begin{equation}
{\cal P}_{\left<^{(L)} h^2 \right>} \sim k^{n-1}_r = k^{3-2\nu}_r.
\end{equation}
Recent calculations \cite{chinos} showed that $n \simeq 1.2$,
which corresponds to $\nu \simeq 1.4$. With this result we obtain
\begin{equation}
m^2  \simeq \left(1.4\right)^2 + \frac{\left[9p(p+1)
-2\right]}{4p^2} ,
\end{equation}
which is the main result of this work. For $p>1$, and $\nu \simeq
1.4$, we obtain the following restrictions for $m$:
\begin{equation}
2.052 < m < 2.44.
\end{equation}

\section{Final remarks}

In this letter we have developed a stochastic approach to study
gravitational waves produced during inflation. We have considered
that the expansion is governed by a decaying cosmological
parameter. The formalism was constructed by considering a 5D
geometrical background which is Riemann flat. Hence, all effective
4D sources are induced from the foliation $\psi=\psi_0$, which is
taken on this 5D flat space-time. In our particular case, the
large scale tensor metric fluctuations are linearized, so that
they obey a like-wave equation of motion. Their components can be
considered on the infrared sector (super Hubble or large scale
tensor fluctuations $^{(L)} h^i\,_{j}$), which obey a set of
stochastic equations, affected by tensor noises $\eta^i\,_{j}$
(gaussian and white in our case, due to the fact we have used a
Heaviside function as a window function on $^{(L)} h^i\,_{j}$).
Due to the isotropy of the 3D space, it is possible to define an
effective diffusion coefficient $D_{\eta\eta}$ to describe the
evolution of $\left<^{(L)} h^2\right>$. In particular, we obtain
that for $\Lambda(t) = 3p^2/t^2$ ($p>1$ and $m=\pm
[1/(2p)]\sqrt{9p(p+1)+4\alpha p-1}$), the parameter $\alpha$ is
restricted by the inequality ${9p^2-1 \over 4 p} = \alpha >0$, for
a scale invariant power spectrum of $\left< ^{(L)} h^2\right>$. In
general, the relevant result of our formalism is that the KK-mass
results to be related to the power of the $\left< ^{(L)}
h^2\right>$-spectrum, $m^2 \simeq n^2 + {\left[9p(p+1)
-2\right]\over 4p^2}$, through the spectral index $n$ and the
parameter $p$ that characterizes the decaying cosmological
parameter $\Lambda(t)$.

\vskip .2cm \centerline{\bf{Acknowledgements}} \vskip .2cm SPGM
acknowledges UNMdP for financial support. MB acknowledges CONICET
and UNMdP (Argentina) for financial support. LFPS acknowledges
CAPES for financial support. JEMA acknowledges CNPq-CLAF
for financial support.
\\

\end{document}